# Nitrogen-vacancy magnetometry of individual Fe-triazole spin crossover nanorods

Suvechhya Lamichhane,[†] Kayleigh A McElveen,[‡] Adam Erickson,[¶] Ilja Fescenko,[§] Shuo Sun,[‡] Rupak Timalsina,[¶] Yinsheng Guo,[‡] Sy-Hwang Liou,[†] Rebecca Y. Lai,[‡] and Abdelghani Laraoui[†,¶,*]

[†]*Department of Physics and Astronomy and the Nebraska Center for Materials and Nanoscience, University of Nebraska-Lincoln, 855 N 16th St, Lincoln, Nebraska 68588, USA*
[‡]*Department of Chemistry, University of Nebraska-Lincoln, 651 Hamilton Hall Lincoln, NE 68588, USA*
[¶]*Department of Mechanical & Materials Engineering, University of Nebraska-Lincoln, 900 N 16th St. W342 NH. Lincoln, NE 68588, USA*
[§]*Laser Center, University of Latvia, Jelgavas St 3, Riga, LV-1004, Latvia*

*Email: alaraoui2@unl.edu

**Abstract**

[Fe(Htrz)$_2$(trz)](BF$_4$) (Fe-triazole) spin crossover molecules show thermal, electrical, and optical switching between high spin (HS) and low spin (LS) states, making them promising candidates for molecular spintronics. The LS and HS transitions originate from the electronic configurations of Fe(II) and are considered to be diamagnetic and paramagnetic respectively. The Fe(II) LS state has six paired electrons in the ground states with no interaction with the magnetic field and a diamagnetic behavior is usually observed. While the bulk magnetic properties of Fe-triazole compounds are widely studied by standard magnetometry techniques their magnetic properties at the individual level are missing. Here we use nitrogen vacancy (NV) based magnetometry to study the magnetic properties of the Fe-triazole LS state of nanoparticle clusters and individual nanorods of size varying from 20 to 1000 nm. Scanning electron microscopy (SEM) and Raman spectroscopy are performed to determine the size of the nanoparticles/nanorods and to confirm their respective spin states. The magnetic field patterns produced by the nanoparticles/nanorods are imaged by NV magnetic microscopy as a function of applied magnetic field (up to 350 mT) and correlated with SEM and Raman. We found that in most of the nanorods the LS state is slightly paramagnetic, possibly originating from the surface oxidation and/or the greater Fe(III) presence along the nanorods' edges. NV measurements on the Fe-triazole LS state nanoparticle clusters revealed both diamagnetic and paramagnetic behavior. Our results highlight the potential of NV quantum sensors to study the magnetic properties of spin crossover molecules and molecular magnets.

Spin crossover (SCO) compounds that experience spin transition phenomena have been studied for decades,[1–8] making them promising candidates in various applications in molecular spintronics,[9] data storage[10,11] and as sensors.[12,13] In particular, SCO iron(II) triazole complex,[14] [Fe(Htrz)$_2$(trz)](BF$_4$)] (Fe-triazole, Htrz = 1H-1,2,4-triazole and trz = deprotonated triazola to (–) ligand) displays two switchable stable spin (low spin, high spin) states that can be controlled by magnetic field,[15,16], optical excitation,[17] temperature[18] even at the individual level.[19] The low spin (LS) and high spin (HS) transitions in the Fe-triazole complex (**Fig. 1a**) originate from the electronic configurations of Fe(II), and are considered to be diamagnetic and paramagnetic respectively.[20–24] The Fe(II) LS state has six paired electrons in the ground states with no interaction with the applied magnetic field and a diamagnetic behavior is usually seen in Fe-triazole complexes at LS.[13,14,22,25]

While the magnetic properties of Fe-triazole compounds in powder or solutions have been studied widely by magnetometry techniques[7,8] such as vibrating-sample magnetometry (VSM), superconducting quantum interference device (SQUID), and Mössbauer spectroscopy, their magnetic properties at the individual level have not been explored. There are recent efforts to study Fe-triazole magnetic properties at the individual level by using for example micro-SQUID magnetomtery.[26] However, the magnetic sensitivity limit for micro-SQUID is not high enough to measure the magnetic moment of individual Fe-triazole nanoparticles.[26] This is related to their weak magnetic moment at the individual level in nanoparticle, nanocrystal, or nanorod configurations;[24] therefore, high sensitive magnetometry techniques with spatial resolution below 400 nm are required.

Recently, a technique for measuring magnetic fields from nanoscale objects has emerged, based on the optical detection of electron spin resonances of nitrogen vacancy (NV) centers in diamond.[27–29] The NV center is a spin-1 defect that features optically addressable electron spin[30] with millisecond quantum coherence at room temperature,[31] and has launched additional frontiers in quantum sensing,[32] nanoscale magnetometry,[33–36] and biosensing.[37] For example, it enabled the first detection of single proteins,[38] perform nanoscale (volume < 1 micrometer cube) nuclear magnetic resonance spectroscopy,[34,39–42] and used very recently to measure the magnetic properties of individual (size < 400 nm) malarial hemozoin biocrystals.[43]

In this paper, we investigate the magnetic properties of individual Fe-triazole nanorods and nanoparticle clusters (size: 20 nm to 1 µm) by using NV magnetometry in combination with scanning electron microscopy (SEM) and Raman spectroscopy to find out the size of the nanorods and to confirm their spin state. We measured spatially resolved magnetic maps of the stray magnetic field produced by the individual Fe-triazole nanorods and nanoparticles clusters as a function of the applied magnetic field (up to 350 mT) and found that most LS states have paramagnetic properties in contrary to early prediction of a diamagnetic behavior in the Fe(II) state.[13,14,22,25]

**Results and discussion**

**Bulk characterization of [Fe(Htrz)$_2$(trz)](BF$_4$) nanoparticle and nanorods powder**
Synthetic protocols were derived from Kroeber *et al.* in reference 25 and Blanco *et al.* in reference 24 to prepare the Fe-triazole nanoparticle (size ~20 nm ± 10 nm) powder and nanorods (length ~ 300 nm – 1 µm) respectively (Methods). We used scanning electron microscopy SEM to determine the lateral dimensions (length and width) of the Fe-triazole nanorods and the size of the nanoparticles. We dispersed the Fe-triazole nanorods/nanoparticles (5 mg/ml) powder on a silicon substrate and performed SEM image (**Fig. S1**) which show a homogeneous distribution of lengths of 1 µm ± 100 nm and widths of 140 nm ± 35 nm across all nanorods, and a diameter of 20 nm ± 10 nm for the nanoparticles. The average height of the nanorods is 75 nm ± 20 nm, determined from atomic force microscopy (AFM) imaging (Methods and **Fig. S3**).

To confirm the LS state of the nanorods (bulk and individual) we used Raman microscopy (Supporting Information Section VI) under ambient condition (1 atm, 298 K). To prevent photoactivation effects induced by the 532-nm laser,[44] we performed Raman measurements on the Fe-triazole nanorods/nanoparticles at different intensities ranging from 1 to 10 mW/µm$^2$ and found no such effects are observed for laser intensity < 5 mW/µm$^2$. **Fig. 1c** shows the Raman spectrum of the 1 µm long Fe-triazole nanorods in the wavenumber range (0-1200 cm$^{-1}$) at laser intensity of 3.5 mW/μm$^2$ with expected features coming from the LS state.[45] When studying the individual nanorods deposited on diamond for NV magnetometry we measured the Raman signal in the wavenumber range 100-340 cm$^{-1}$ (insert of **Fig. 1c**) as the diamond substrate produces a high Raman signal for wavenumbers > 400 cm$^{-1}$ [46] that hinders any Fe-triazole signal.

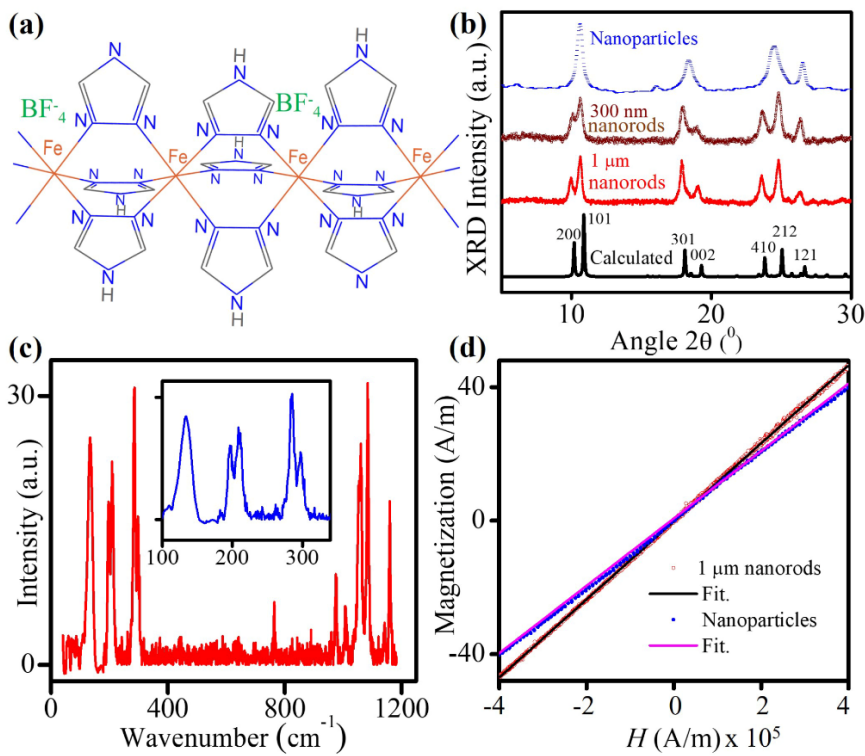

**Figure 1. (a)** Molecular Structure of [Fe(Htrz)$_2$(trz)](BF$_4$) spin crossover complex derived from reference [25]. **(b)** Measured XRD spectra of the Fe-triazole 20 nm ± 10 nm nanoparticles (blue dash scattered line), 300 nm rods (brown open circle scattered line), and 1 μm rods (red closed circle line) Fe-triazole nanorods. The measured data spectra are compared with calculated spectrum (black solid line) of [Fe(Htrz)$_2$(trz)](BF$_4$)]. **(c)** Normalized Raman spectrum of the LS state of the 1 μm Fe-triazole nanorod powder measured at room temperature. Insert of **(c):** a zoomed Raman spectrum in the wavenumber range of 100-330 cm$^{-1}$ to compare with individual measurements done on the diamond (**Fig. S7**). **(d)** *M-H* curve of the 1 μm Fe-triazole nanorods and nanoparticles powder measured at 295 K.

To estimate the density of the Fe-triazole nanoparticles and nanorods we performed X ray diffraction (XRD) spectroscopy at room temperature by using a Bruker-AXS D8 Discover diffractometer with Cu-Kα x-ray, Ge monochromator and Vantec 500 area detector. **Figure 1b** shows the measured XRD spectra recorded from Fe-triazole 20 nm nanoparticles, 300 nm and 1 μm nanorods, compared with calculated spectrum retrieved from The Cambridge Crystallographic Data Centre (www.ccdc.cam.ac.uk/data_ request/cif).[20] The calculated XRD peaks overlap well with the measured peaks in the nanorods (300 nm, 1 μm) with a broadening, explained by the variation of the nanorods' grain size. However, in nanoparticles, there is a slight change. For example, the measured XRD peak at 2θ =10.6$^0$ presents a singlet while the calculated one is a doublet, *i.e.*, two peaks appear at 2θ = 10.18$^0$ and 10.88$^0$ respectively. We explain this variation by the existence of [Fe(Htrz)$_2$(trz)](BF4) as two polymorphs I and II.[47] The Fe-triazole nanoparticles are made using a direct synthetic approach with no surfactants.[25] This method usually results in Polymorph II and the XRD spectrum has a fewer and broader peaks in comparison to the calculated ones (see **Fig. S2**). The Fe-Triazole nanorods however are made by the reverse micellar approach that utilizes a surfactant (Tergitol NP9) and water as the solvent,[24] which results in Polymorph I, and the XRD spectrum has multiple and narrower peaks (similar to **Fig. 1b**).[48] The broadening of XRD peaks in Fe-triazole nanoparticles is explained by the variation of their size. By using the lattice parameter inferred from XRD measurements in **Figs. 1b**, **S2** and the molecular weight of Fe-triazole (monomer) of 348.8 g/mol[48], we calculated the Fe-triazole density of 1.9847 g/cm$^3$ ± 0.12 g/cm$^3$ for the nanoparticles and 1.8515 g/cm$^3$ ± 0.06 g/cm$^3$ for the 300 nm and 1 μm nanorods respectively.

To confirm the temperature dependent spin states (LS and HS) of the Fe-triazole nanorods we used VSM (Versa Lab 3T, Quantum Design) (**Fig. S4**). All the measurements are performed with samples undergoing a second heating and cooling cycle to allow for residual solvent evaporation (see **Fig. S5**). In this paper we focus mainly on the magnetic properties of the LS state of the Fe-triazole nanoparticles and nanorods. **Figure. 1d** shows the magnetization-magnetic field (*M-H*) curves of the Fe-triazole 1 μm nanorods' and nanoparticles' powder, measured at 295 K, and confirms a slight paramagnetic behavior with a low magnetization of 40 A/m at applied field of 3.49 x 10$^5$ A/m (438 mT) and 3.94 x 10$^5$ A/m (~ 494.7 mT) respectively. Magnetization is calculated by multiplying the magnetic moment per mass (emu/g) from VSM measurements with the Fe-triazole density computed from XRD measurements (**Fig. 1b**). A similar behavior is observed in Fe-triazole 40-50 nm nanocube powder, and explained by the formation of Fe(III) at the surface of the cubes.[24] The magnetic susceptibility ($\chi$) ~ 1.17 x 10$^{-4}$ ± 1.75 x10$^{-5}$ for the 300 nm and 1 μm Fe-triazole nanorods' powder, and 1.01 x 10$^{-4}$ ± 1.51 x 10$^{-5}$ for the nanoparticles' powder. Bulk VSM measurements yield ensemble-averaged properties, and it is difficult to extract individual measurements from strong magnetic interactions or contamination during preparation. To confirm whether the paramagnetic behavior persists at the individual level we performed NV magnetometry measurements of Fe-triazole nanorods and nanoparticles (discussion below).

**NV magnetometry detection protocol**

The negatively charged NV center, comprised of a substitutional nitrogen adjacent to a vacancy site (**Fig. 2a**), is a photostable emitter that displays a high (up to 30%) optically detected magnetic resonance (ODMR) contrast.[30] The NV spin-triplet ground ($^3A_2$) state features a zero-field splitting $D$ = 2.87 GHz between the electronic spin states $M_s$ = 0 and $M_s$ = ±1 (**Fig. 2b**). Laser illumination (500-630 nm) produces spin-conserving excitation to the excited triplet state ($^3E$), which in turn leads to far-red fluorescence (650-800 nm). Intersystem crossing to metastable singlet states takes place preferentially for NV centers in the $M_s$ = ±1 states, ultimately resulting in an almost-complete transfer of population to the $M_s$ = 0 state; this same mechanism allows for optical readout of the spin state *via* spin-dependent fluorescence.[30] The application of an external magnetic field $B_{app}$ breaks the degeneracy of the $M_s$ = ±1 state and leads to a pair of spin transitions $|M_s = 0 >$ to $|M_s = -1 >$ (referred as $f_-$) and $|M_s = 0 >$ to $|M_s = +1 >$ (referred as $f_+$) whose spin resonance frequencies depend on the magnetic field component along the NV symmetry axis (**Fig. 2c**).

In **Figure 2d** we depict the widefield ODMR microscope used for imaging individual Fe-triazole nanorods and nanoparticle clusters. We used a 532-nm laser (180 mW) to excite the NVs over an area of ~ 39 x 39 μm$^2$ and the NV fluorescence (650-800 nm) is mapped onto a sCMOS camera. We detail the experimental setup in Methods. For the NV magnetometry measurements reported in this study, we used 2 mm x 1 mm x 0.1 mm type-Ib high pressure high temperature (HPHT) (110) cut and polished diamond. The 200 nm thick NV sensing layer is created near the diamond surface by using $^4$He$^+$ implantation followed by ultravacuum (down to 10$^{-6}$ Torr) high temperature (1373 K) annealing and cleaning in a boiling tri-acid mixture (Methods).[41,43,49] The diamond substrate with Fe-triazole nanorods/nanoparticles is placed on top of a glass coverslip with patterned gold loops for microwave (MW) excitation. The Fe-triazole nanorods are magnetized under an external magnetic field $B_{app}$, generated by two permanent magnets, oriented along the NV [111] axis (parallel to $x$) in (110) diamond. The Hamiltonian of the ground spin state of NVs aligned along [111] direction ($x$, **Fig 2d**)) is:

$$H \cong DS_x^2 - \gamma_{NV}(S_x(B_{app} + B_{str})),$$

where $\gamma_{NV} = 28$ GHz/T is the gyromagnetic ratio of the electronic spin. The magnetized Fe-triazole nanorods produce a stray magnetic field $B_{str}$ that leads to a change in the effective field $B_{NV} = B_{app} + B_{str}$ and to corresponding changes of the frequencies of the NV electronic spin transitions $f_+$ and $f_-$. By sweeping the MW frequency across the $f_-$ transition (**Fig. 2e**) and $f_+$ transition (**Fig. 2f**) we record the ODMR peaks. Then by fitting them with a Lorentzian, we determine $B_{str}$ generated from the Fe-triazole nanorods and nanoparticles. This is the basis of DC magnetic sensing with NV centers.[28,30,32,34–36,43]

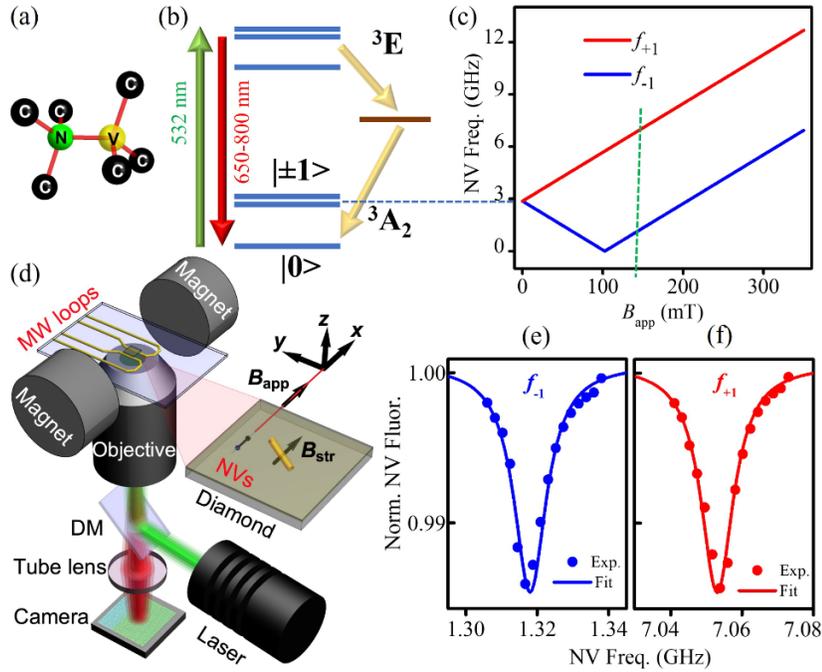

**Figure 2.** (a) A schematic of the NV center (nitrogen: green atom, yellow: vacancy) in the diamond lattice. (b) A schematic of the energy levels of the NV center ground ($^3A_2$) and excited ($^3E$) states with intermediate metastable state. A green laser (532 nm) initializes the NV center spin and results in fluorescence in wavelength range of 650-800 nm. (c) Calculated field dependence of NV spin transitions $f_-$ and $f_+$ frequency as function of the applied magnetic field. (d) A schematic of the ODMR microscope in wide-field geometry (see **Fig. S6c** for the setup picture) used to study the Fe-triazole nanorods and nanoparticle clusters. DM stands for a dichroic mirror. A magnetic field $B_{app}$ is applied along $x$ for (110) diamond. ODMR peaks of NV spin transitions $f_-$ (e) and $f_+$ (f) recorded at $B_{app}$ = 149.5 mT.

NV magnetic maps are obtained by performing ODMR measurements, *i.e.*, sweeping the MW frequency across $f_-$ and $f_+$ spin transitions and recording the fluorescence images at each MW frequency step by the sCMOS camera.[43] The ODMR signal is fitted for each pixel by a Lorentzian, like in **Fig. 2e** and **Fig. 2f**, to deduce the NV spin resonance frequencies $f_-$ and $f_+$. We then calculated the effective field $B_{NV} = B_{app} + B_{str} = (f_+ - f_-)/(2\gamma_{NV})$, and subtracted $<B_{NV}> \approx B_{app}$ from it. We sum up $f_+$ and $f_-$ frequencies because at $B_{app}$ = 149.5 mT frequency $f_-$ is on the other side of the ground-state level anti-crossing, (see **Fig. 2c**). The resulting magnetic-field map reveals only the stray field $B_{str}$ produced by the Fe-triazole nanorods/nanoparticles deposited on top the diamond (NV-layer side). This measurement process removes any artifact signal coming from contamination or residue of the Fe-triazole/Ethanol solution, temperature dependent changes of the zero-field splitting,[50] and strain variation within the diamond substrate.[51]

The minimum measurable magnetic field in the photon-shot-noise limit is:[36,43]

$$B_{min} \cong 4\,\Gamma\,(3\sqrt{3}\,\gamma_{NV}\,C)^{-1}\,(I_0\,t)^{-1/2},$$

Where $C$ is the ODMR contrast (= 0.014), $\Gamma$ is the full-width-at-half-maximum (FWHM) linewidth of the ODMR peak (= 11 MHz), $I_0$ is the NV fluorescence intensity (= 1.74 $10^8$ c/s for a detection voxel of 390 x 390 $nm^2$), and $t$ is the averaging time.[52] By using the measurements parameters in **Fig. 2e** and **Fig. 2f** we estimate $B_{min} \simeq 1.4$ μT for $t$ = 1s. By increasing the averaging time ($\geq$ 600 s) $B_{min}$ values of 0.3 μT can be reached (15% close to the shot noise limit).[43] This is sufficient to image individual Fe-triazole nanorods and nanoparticle clusters.

**NV magnetometry of the low spin state of individual 1 μm Fe-triazole nanorods**
To perform NV measurements, we diluted 2 μL of the 1 μm Fe-triazole suspension (5 mg/mL) in 2 mL of ethanol and sonicated it for 15 minutes to prevent formation of clusters. We then drop cased 10 μL of the Fe-triazole/ethanol solution on the diamond substrate and let it to dry for a few minutes. **Figure 3a.** shows the SEM image of the 1 μm Fe-triazole nanorods on top of the diamond, patterned with references letters by Focused Ion Beam (FIB) for tracking (Methods). Most of the nanorods are single, with only a few clusters, and produce a magnetic pattern in the NV ODMR image in **Fig. 3b**. To confirm whether these Fe-triazole nanorods preserved the LS state after SEM and NV measurements, we performed Raman spectroscopy on selected nanorods and correlated with SEM and NV measurements (see **Figure S7** in Supporting Information Section VI). Indeed, all nanorods that produce a magnetic pattern in the NV image exhibit LS state as expected from Fe-triazole complexes.[14,20,21,24,25,53]

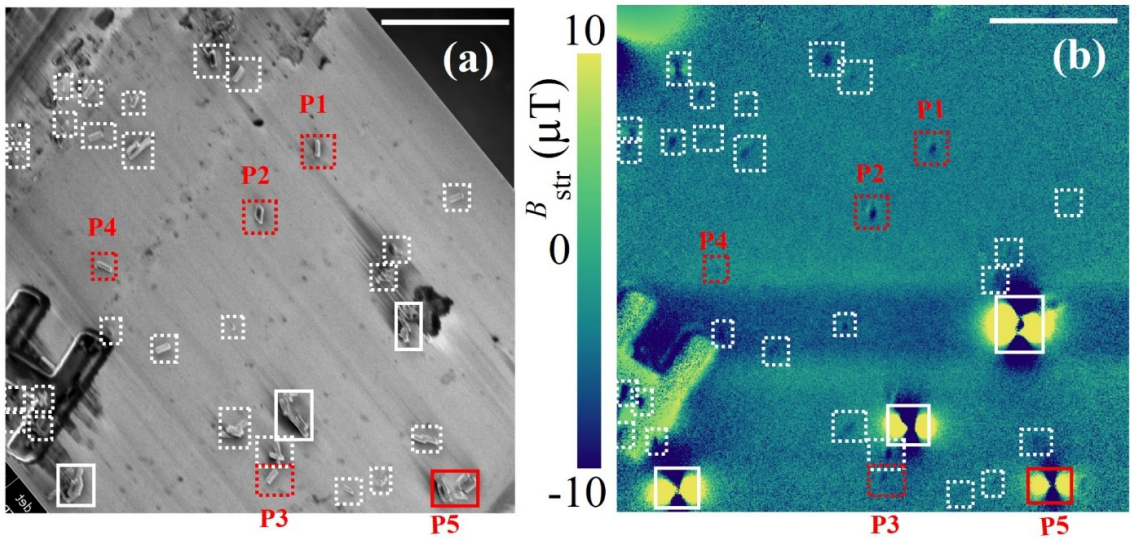

**Figure 3. (a)** SEM image of the 1 μm Fe-triazole nanorods drop-casted on the diamond surface marked with reference letters by focused ion beam. **(b)** NV ODMR map of the same region in **(a)** for $B_{app}$= 85 mT. The labeled rods with red numbers are shown in detail in **Fig. 4**. Scale bare is 10 μm in **(a)** and **(b)**.

**Figure 3b** displays the NV produced magnetic-field map of the Fe-triazole 1 µm nanorods in **Fig. 3a**. at an applied magnetic field of 85 mT. Of the 33 nanorods in **Fig. 3b**, 29 nanorods (dashed squares) produce weak magnetic stray field (≤ 10 µT), and 4 rods (solid squares) generate strong magnetic features ($B_{str}$ > 10 µT). To further investigate the magnetic properties of the LS state at the individual level, we labelled selected Fe-triazole nanorods with numbers (P1-P5) in **Fig. 3** and modeled their magnetic stray fields considering their dimensions inferred from SEM and AFM (discussion below).

To accurately describe the diamond magnetic microscope response, a magnetostatic finite-element simulation is used to calculate the magnetic field $B_{str}$ (x, y, z) produced by elongated ellipsoids with uniform magnetization (Supporting Information Section VII) with lateral dimensions and thickness taken from SEM images in **Fig. 4a** and AFM images in **Fig. S3** respectively. We then integrated $B_{str}$ over the NV vertical (z) distribution, which is assumed to be uniform from ~ 2 to 202 nm below the diamond surface (see Methods and Supporting Information Section V). **Figure 4c** shows the modeled magnetic field patterns for selected Fe-triazole nanorods (P1-P5) and described with a volume magnetization $M$ = 40 A/m that corresponds to a volume susceptibility $\chi = \mu_0 M/B_{app}$ = 1.58 x $10^{-4}$, where $\mu_0$ is the vacuum permeability = $4\pi \times 10^{-7}$ m T/A. The calculated susceptibility value agrees well with the value obtained from bulk VSM (**Fig. 1d**) and SQUID measurements reported in reference 24. The nanorods (dashed squares in **Fig. 3**) produce weak magnetic-field pattern with a variation of $\chi$ in the range of 48% we explain the reasons below and in Supporting Information Section X. The used model does a good job in reproducing the measured patterns in **Fig. 4b**, except from the nanorods highlighted by solid squares (e,g, P5) in **Fig. 3**, which have an amplitude 10 times lower than the measured ones, suggesting a different magnetic behavior.

To further elucidate their magnetic properties, we performed line cuts of the measured and calculated magnetic field patterns by averaging the $B_{str}$ values over six rows (390 nm) in the region highlighted by dashed lines in **Figs. 4b** and **4c** respectively. We then calculated the amplitude $\Delta B$ by subtracting the maximum and minimum values of $B_{str}$, **Fig. 4d**. $\Delta B$ varies in the range of 1 - 7 µT ± 0.2 µT based on the nanorod shape, size, and alignment along the applied magnetic field $B_{app}$. This is clearly seen for nanorod P4, aligned parallel to $B_{app}$, where $\Delta B$ is ~ 1 µT below the value of 3 µT obtained in nanorod P1, aligned perpendicular with $B_{app}$. The value of $\Delta B$ in nanorod P1 (thinner with width of 250 nm) is 3.2 µT below the measured value of 6.4 µT in nanorod P2 (thicker with width of 500 nm). The amplitude $\Delta B$ for cluster P5 is 22.7 µT ± 0.2 µT with a spatial FWHM ~ 475 nm of the magnetic pattern. Similar values of $\Delta B$ and FWHM are obtained on three clusters (highlighted by solid squares in **Fig. 3b**), and we describe below the possible origins.

To check the magnetic behavior of the studied Fe-triazole 1 µm nanorods we performed ODMR maps, similar to **Fig. 3b**, at different values of the applied magnetic field $B_{app}$ in the range of 2 – 350 mT. **Figure 4e** displays the magnetic pattern amplitude $\Delta B$ of the nanorods P1-P5 as function of $B_{app}$. Linear dependence is observed in nanorods P1-P4 (**Fig. 4e**) and on other 29 rods (highlighted with dashed squares in **Fig. 3**) which is typical of paramagnetic behavior. We used a linear function ($\Delta B = A \, B_{app}$) to fit the measurements (solid lines in **Fig. 4e**), where $A$ is the curve slope, and it corresponds to the magnetic susceptibility $\chi$ of the nanorods. Susceptibility $\chi$ varies in the range from 1.65 x $10^{-4}$ to 3.2 x $10^{-4}$ in the 29 rods (**Fig. S12**), which corresponds to a variation of 48%. The uncertainty in estimating $\chi$ can be explained by the NV-layer imperfect distribution, image blur, nanorod dimensions, orientation, and shape with leads to lateral anisotropic components of the magnetization along the NV axes.[43] In our model (Supporting Information

Section VII) we considered the nanorods as elongated ellipsoids with a uniform isotropic medium. In ferromagnetic nanoparticles and nanodots with a strong magnetic anisotropy, the shape strongly affects their magnetic properties.[54–57] We discuss the χ variation for Fe-triazole nanoparticles, 300 nm and 1 μm nanorods in the Supporting Information Section IX.

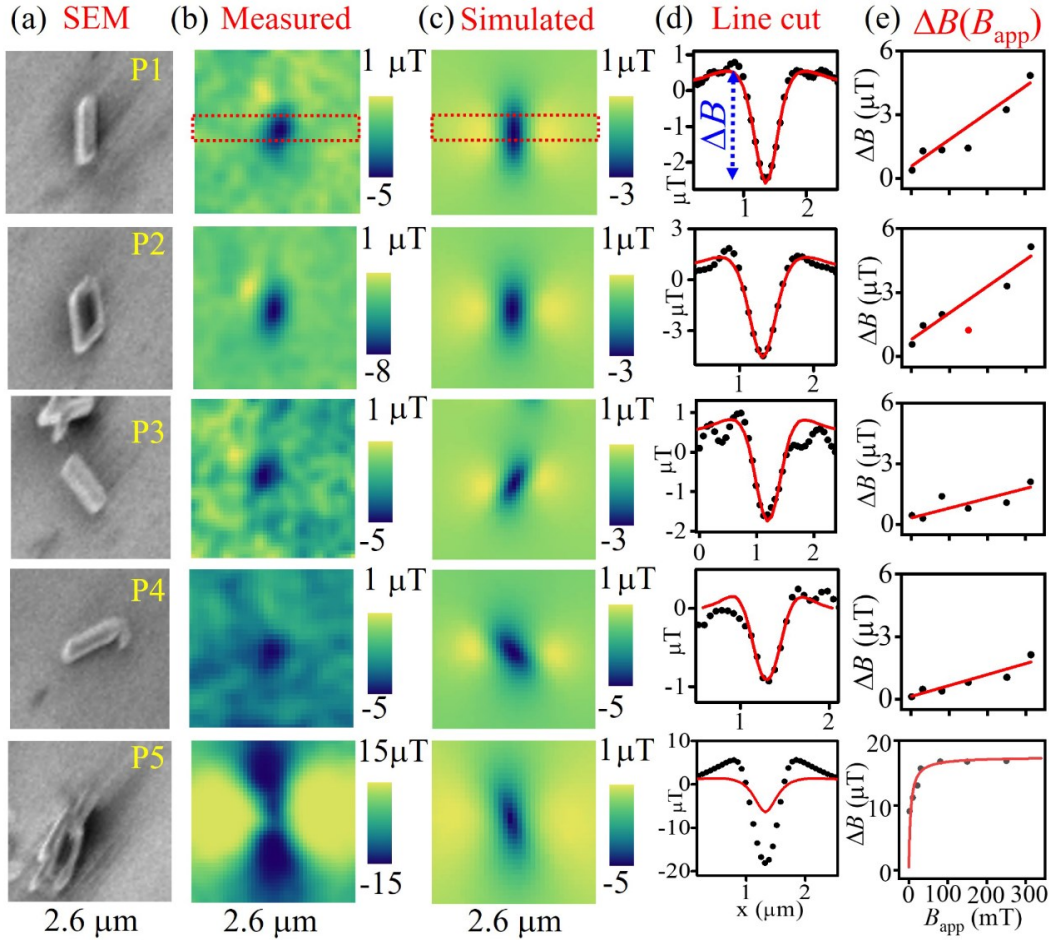

**Figure 4. NV magnetic imaging of individual 1 μm Fe triazole nanorods. (a)** Zoomed SEM images of the Fe-triazole nanorods labeled P1-P5 in **Fig. 3a**. **(b)** Corresponding NV ODMR magnetic maps at $B_{app}$= 85 mT for each nanorod. **(c)** Modeled magnetic-field maps obtained with nanorod sizes deduced from SEM in **(a)** and AFM in **Fig. S3**. We used magnetic susceptibility χ of ~ 1.58 x 10$^{-4}$ at $B_{app}$= 85 mT. **(d)** Line cuts for studied magnetic patterns (scattered closed circle: NV measured, line: modeled), from which the stray field amplitude Δ$B$ is deduced. **(e)** Δ$B$ dependences on the $B_{app}$ for studied nanorods (scattered closed circles: NV measurements, solid lines: fits). P1-P4 are paramagnetic based on the linear fit (similar to the Fe triazole rod powder in **Fig. 1d**). P5 exhibits a superparamagnetic behavior and is fitted with a Langevin showing a saturation response at ~ 100 mT.

For nanorod P5, Δ$B$ increases nonlinearly with applied magnetic field and saturates at $B_{app}$ ~ 100 mT. This is typical superparamagnetic behavior and the measured Δ$B(B_{app})$ is fitted well with a Langevin.[58] This is observed only on 4 (highlighted by solid squares in **Fig. 3c**) of the 33 studied nanorods, highlighted with solid squares in **Fig. 3**, and their magnetic field patterns are aligned in the same way as ones of the paramagnetic nanorods. Usually, superparamagnetic behavior appears only in small ferromagnetic or ferrimagnetic nanoparticles where the thermal energy exceeds the magnetic anisotropy.[58] In our NV measurements we observed this behavior in some of the clustered rods, not individual ones, suggesting the possible formation and trapping of iron oxide nanoparticles between the Fe-triazole rods.[59]

**NV magnetometry of the low spin state of Fe-triazole nanoparticles clusters**

Most of the previously studied bulk Fe-triazole systems are pseudo-octahedral $3d^6$ iron(II) complexes, which LS exhibits a diamagnetic behavior (S = 0).[14,20,22,23,25,60] A relatively high magnetic moment, compared to bulk films, was measured in the LS state in Fe-triazole nanoparticles,[13,47,61] and attributed to their high surface area.[61] The surface Fe atoms have a partial coordination, preferring their HS state at all temperatures.[61] To confirm whether this is still the case in our prepared Fe-triazole nanoparticles (size: 20 nm ± 10 nm) we performed NV magnetometry measurements. We drop-casted the Fe-triazole 20 nm nanoparticles on top of the diamond substrate and performed ODMR imaging, **Fig. 5a**. While sonicating the diluted Fe-triazole/ethanol solution to prevent clustering (Methods), we still observe nanoparticle clusters as confirmed by optical image (**Fig. S9a**) and SEM (**Fig. S1a**).

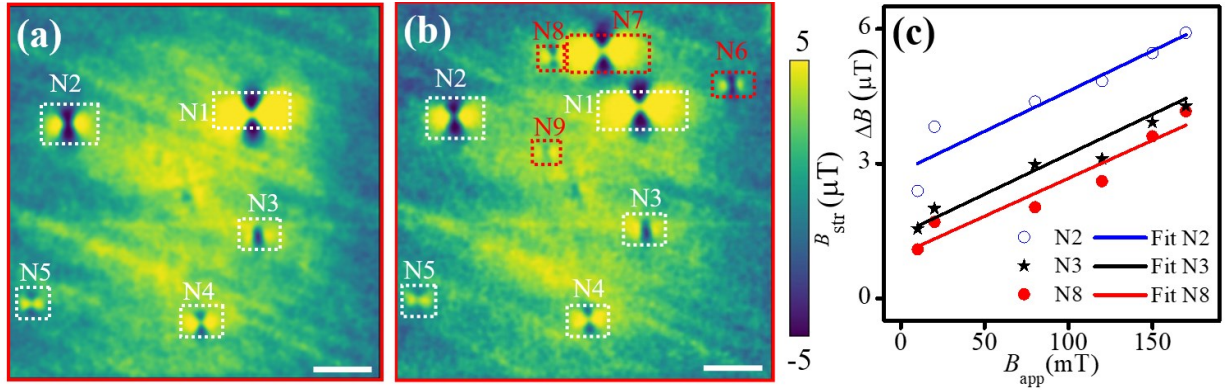

**Figure 5. NV magnetic imaging of Fe-triazole nanoparticles.** NV magnetic-field maps generated from the Fe-triazole nanoparticles (20 nm ± 10 nm) deposited on top of the diamond at 295 K **(a)** and after heating to 395 K and then cooling to 295 K **(b)**. **(c)** Measured (scattered) magnetic pattern amplitude $\Delta B$ as function of the applied field for selected particles in **(c)** fitted (solid lines) with linear function, confirming the paramagnetic behavior. The scale bar in **(a)** and **(b)** is 5 μm.

**Figure 5b** shows the ODMR map recorded from the Fe-triazole nanoparticle clusters at 295 K. Not all nanoparticle clusters produce magnetic patterns as expected from the diamagnetic LS state. However, five clusters (N1-N5) produce a magnetic pattern with a stray field $B_{str}$ in the range of -5 μT to 5 μT at $B_{app}$ = 250 mT. We then heated the diamond substrate at 395 K around the HS temperature transition (**Fig. S4a**), cooled down to 295 K, and performed NV ODMR imaging (**Fig. 5b**) of the same region in **Fig. 5a**. Additional magnetic patterns appeared (N6-N9) with similar stray field values. Repetitive NV measurements at 295 K (**Fig. S9b**) show that the appearing patterns (N6-N9) originate from the LS state with a possible residual fraction of the HS state. Due to the significant thermal drift at high temperatures (> 400 K) we did not measure the HS of the nanoparticles as in **Fig. 5b**. These measurements may indicate the possible formation of Fe(III) oxide on the nanoparticles' surface[24] upon heating, which is paramagnetic, rather than a switching between the LS and HS states. The pattern amplitude $\Delta B$ dependence with applied magnetics $B_{app}$ for N1-N9 nanoparticle clusters (**Fig. S13**) confirm the paramagnetic behavior with a magnetic susceptibility of $1.8 \times 10^{-4} \pm 2.95 \times 10^{-5}$. **Fig. 5c** shows $\Delta B$-$B_{app}$ curves for selected Fe-triazole nanoparticles (N2, N3, and N8).

**Conclusions**

In summary, we used diamond magnetometry to study the low spin state magnetic properties of Fe triazole nanoparticles and individual nanorods of size varying from 20 nm to 1000 nm. Most of the 300 nm and 1 μm nanorods display paramagnetic behavior, with magnetic field patterns well defined by a magnetostatic model using a volume susceptibility $\chi \sim 1.58 \times 10^{-4}$, that may be explained by the formation of Fe(III) at the nanorods' surface.[24] The rest of the studied nanorods exhibit a large magnetization that saturates at magnetic fields $\sim$ 100 mT, suggesting a superparamagnetic behavior. This behavior is observed only in some of the clustered rods, not individual ones, suggesting the possible formation and trapping of iron oxide nanoparticles between the Fe-triazole rods during synthesis.[59] Further work is needed to determine the composition and structure of the nanorods surface by using X-ray absorption spectroscopy (XAS) and X-ray photoemission spectroscopy (XPS),[62] TEM microscopy,[13] and Lorentz microscopy.[63]

NV measurements on Fe-triazole nanoparticle clusters revealed both diamagnetic and paramagnetic behavior of the LS state at 295 K. Repeated measurements at 295 K (LS) after heating at 395 K (around the HS transition) suggested the formation of Fe(III) oxide on the nanoparticles' surface, similar to earlier ensemble measurements performed on Fe-triazole nanocubes.[24] The difficulty in correlating NV magnetic images with SEM images due to charging issues,[64] makes it hard measuring on individual Fe-triazole nanoparticles. Coating Fe-triazole nanoparticles on top of diamond with a conductive polymer can enable imaging below 20 nm, as demonstrated in iron oxide nanoparticles.[65] Decreasing the size of the Fe-triazole nanoparticles leads to a decrease of the produced stray fields < 0.1 μT and make the NV CW measurements less sensitive (the sensitivity of our setup is $\sim$ 1.4 mT. $Hz^{-1/2}$). With improvements in the experimental setup, for example, by doping the diamond with high density of NVs,[66] and performing pulsed NV measurements[67] could improve the sensitivity to below 50 nT $Hz^{-1/2}$ and study the magnetic properties of individual nanoparticles below 10 nm.

**Methods**

**Synthesis of [Fe(Htrz)$_2$(trz)](BF$_4$) nanoparticle and nanorods powder**

For the nanoparticle powder we used the recipe derived from reference 25. A 3 mol solution of 1,2,4-triazole in anhydrous ethanol was added while stirring to a solution of 0.5 mol Fe(BF$_4$)$_2$ · 6 H$_2$O in anhydrous ethanol at 30 μL/min. The resulting pink suspension was stirred for one hour and then left overnight. The solid was then washed thoroughly with ethanol. We did transmission electron microscopy (TEM) measurements (not reported here) and found an average Fe-triazole nanoparticle size of 19 nm ±7 nm.

For the nanorods we followed the recipe developed by Blanco et al. in reference 24. Micellar solution 1 was prepared by combining 0.360 g [Fe(BF$_4$)] · 6H$_2$O (Aldrich) and 10 mg ascorbic acid (Sigma-Aldrich) in 1 mL water and stirring until dissolved. Micellar solution two is prepared by placing 0.220 g 1,2,4-triazole (Alfa Aesar) in 1 mL water. Once both micellar solutions are dissolved, 4 g NP-9 Tergitol surfactant (Sigma-Aldrich) is added to each, and both are stirred separately at 353 K for 5 minutes and 10 minutes for the 300 nm and 1 μm rods respectively. Both solutions are quickly combined and are left stirring for 24 hours at room temperature. The reaction is stopped using ethyl acetate, and then is centrifuged at 4500 rpm for 30 minutes three times each using ethyl acetate as the solvent. The final product is gravity filtered and rinsed thoroughly with water and ethyl acetate until dry.

**SEM measurements of Fe-triazole nanoparticles and nanorods**

To measure the size and dimensions of the Fe-triazole nanorods we dispersed the 300 nm and 1 μm rods powder on a silicon substrate and performed SEM imaging (FEI Helios NanoLab 660, voltage = 5 kV). For the Fe-triazole nanoparticles we drop-casted them directly in a carbon tape to prevent the charging issue during SEM measurement. **Fig. S1a** displays the SEM image obtained on the nanoparticles with a size of 20 ± 10 nm. **Fig. S1b** (**Fig. S1c**) show SEM image of the 300 nm (1 μm) nanorods with a distribution of lengths of 300 nm ± 50 nm (1 μm ± 100 nm) and widths of 80 nm ± 20 nm (140 nm ± 35 nm) respectively.

**AFM height measurements**

To estimate the height of the Fe-triazole nanorods, we used an atomic force microscopy (AFM, Digital instruments 3000). We diluted 2 μL of the 1 μm Fe-triazole solution (5 mg/mL) in 2 mL of ethanol and sonicated it for 15 minutes to prevent clustering. We then drop-casted 10 mL of the Fe-triazole/ethanol solution on a Si substrate and let it dry for a few minutes. **Figure S3a** displays the AFM topography image of the 1 μm Fe-triazole rods and their height is estimated by drawing the profile line across each individual nanorod. The height distribution is shown in **Fig. S3b** with a mean height of ~ 75 nm ± 20 nm. We inferred the mean height of the nanorods in the magnetostatic simulations to calculate the stray-field produced from each nanorod (Supporting Information Section VII).

**Creation of NV centers in diamond**

We used 2 mm x 2 mm x 0.5 mm type-Ib high pressure high temperature (HPHT) (100) diamond (Sumitomo) substrate with a nitrogen concentration of ~ 100 ppm. The diamond is cut and polished along (110) plane at Delaware Diamond Knives Inc to 2 mm x 1 mm x 0.08 mm membranes. The diamond is then implanted at CuttingEdge Ions LLC with $^4He^+$ ions at three energies (5, 15, and 33 keV) and three doses (2. $10^{12}$, 2. $10^{12}$, and 4. $10^{12}$ cm$^{-2}$) respectively to create a uniform layer of vacancies near the diamond surface.[43,49] We used Stopping and Range of Ions in Matter (SRIM) Monte Carlo simulations to estimate the vacancy distribution depth profile in the diamond substrate (**Fig. S6a**) and found a uniform distribution within ~ 200 nm beneath the diamond surface facing the $^4He^+$ source. After the implantation we annealed the diamond substrate in an ultravacuum (pressure $\leq 10^{-6}$ torr) furnace at 1073 K for 4 hours and at 1373 K for 2 hours,[43] and then cleaned it for two hours in a 1:1:1 mixture of nitric, sulfuric, and perchloric acid at 473 K to remove graphite reside at the suraface.[41,43,49] This process resulted in ~ 200-nm NV layer near the surface with a density of ~ 10 ppm, based on the fluorescence measurements in **Fig. S6b**.[43,68]

**Optical detected magnetic resonance setup**

The wide field microscope (WFM, **Fig. S6c**) used in this study is a home-built optical detected magnetic resonance (ODMR) setup combined with sCMOS camera (upper insert of **Fig. S6c**) for wide-field imaging.[35] We used a long working distance (1 mm), high NA (= 0.9) 100x Nikon objective to focus the green laser (532 nm) with a variable laser power (up to 2 W). The diamond substrate with Fe-triazole nanorods (lower insert of **Fig. S6c**) is attached to a glass coverslip (thickness of 100 μm) with two microwave (MW) Au loops patterned on it for NV spin excitation. The coverslip with diamond is then mounted on a XYZ motorized stage (Newport, Picometer 8742) that is connected to the PC through USB interface for imaging a selected region of Fe-triazole nanorods. The fluorescence (650 – 800 nm) is collected by the same objective, transmitted through a dichroic mirror (DM, Semrock model #FF560-FDi01) and a single-band bandpass filter (Semrock FF01-731-137), and then focused on the sCMOS camera (Hamamatsu, ORCA-Flash4.0 V3) via a tube lens (Thorlabs, TTL200). For magnetic field alignment the NV fluorescence is reflected by a flip mirror and focused with a lens (focal length of 30 mm) on an avalanche photon detector (APD, Thorlabs APD410A), connected to a Yokogawa oscilloscope (DL9041L).

The MW is provided by a signal generator (SRS: Stanford Research Systems, SG384) with two outputs for the first (DC-4 GHz) and double-frequency harmonic (4-8 GHz). It is then connected to two MW amplifiers (Minicircuits ZHL-16W-43-s+ and ZVE-3W-83+) and sent through the blue SMA cables in **Fig. S6c**. For magnetic fields > 185 mT ($f_+$ > 8 GHz) we used additional frequency doublers (Mini-Circuits ZX90-2-24-S+ and ZX90-2-36-S+) on both SRS harmonic outputs to access higher microwave frequencies (> 8-15 GHz), that are amplified by a MW amplifier ZVE3W-183+ (Minicircuits). The MW frequency is swept across the $|m_s = 0>$ to $|m_s = -1>$ ($f_-$) and $|m_s = 0>$ to $|m_s = +1>$ ($f_+$) NV spin transitions by sending a ramp function to the SRS generator analog frequency-modulation input. An output modulation ramp pulse is then sent to the oscilloscope and is triggered with the APD (NV fluorescence) signal to perform live ODMR measurements. Two permanent magnets (KJ magnetics, DX8X8-N52) are connected through stainless steel nonmagnetic structures for better stability and provide a homogeneous magnetic field in the range of 0 to 350 mT. There are four sub-ensembles of NV centers with different symmetry axes and in this study,[49,50] we used only the NV spins oriented along the [111] direction in (110) diamond for measuring the stray field generated by the Fe-triazole nanorods. By monitoring the ODMR peaks in the oscilloscope along the [111] ($x$) direction we can align the applied magnetic field very accurately. After the magnetic field alignment, the mirror is flipped down and the NVs fluorescence is imaged by the sCMOS camera, and each MW frequency is swept along $f_-$ and $f_+$ NV spin transitions. More details for such measurements are provided in reference 43.

**Data acquisition**
We used the following parameters for the camera: frame size is 600 pixels x 600 pixels (1 pixel = 65 nm at 100x magnification), the field of view is 39 μm x 39 μm, and the exposure time is 5 ms. An ODMR image of 39 μm x 39 μm region (*e.g.*, **Fig. 3b**) takes a few minutes to tens of minutes depending on the integration time (3 to 10 ms) of the sCMOS camera and the number (16- 40) of MW frequencies swept across the ODMR peaks. To correct the mechanical and thermal drifts occurring during the NV measurements we patterned the diamond substrate with micron-scale letters (**Fig. S6d**) by etching features in the NV layer. We used Focused Ion Beam (FIB, provided by FEI Helios NanoLab 660) with a voltage of 30 kV and a current of 0.79 nA to etch 500 nm down the diamond film. The etched dark regions in **Fig. S6b** (*e.g.*, $x$ and $H$) have no fluorescence (no NVs) and by using LabVIEW intensity tracking we can digitally correct from any drift and have a very good stability for hours (drift < 100 nm/h).

**Associated Content**

**Supporting Information**
SEM measurements; XRD analysis; AFM measurements to confirm the dimension and height of the nanoparticles/nanorods; magnetic thermal properties of Fe Triazole nanoparticle/nanorod powder; experimental details: Creation of nitrogen vacancy (NV) centers in diamond, and optical detected magnetic resonance setup; Raman and NV magnetometry characterization of the LS state of the Fe-triazole 1 μm nanorods; calculation of the magnetic stray field generated from the Fe-triazole nanorods; additional NV measurements on Fe-triazole 20 nm nanoparticles; uncertainty in susceptibility measurements; and summary of $\Delta B$-$B_{app}$ measurements on Fe-triazole nanoparticles/nanorods.

**Author Contributions**
S.L. and A.L. performed NV measurements; K.A.M synthesized the Fe-triazole powders (nanoparticles and nanorods). S.L. and A.E. took the SEM images; I.F. performed the finite-elements simulations and wrote the Mathematica code to fit the measured magnetic patterns of the Fe-triazole nanorods; S.S. and Y.G. did Raman measurements on the Fe-triazole nanorods; R.T. helped S.L. with the optical alignment of the ODMR setup and in diamond annealing; S.-H.L, R.Y.L., and A.L. designed the experiments and supervised the project. A.L. wrote the manuscript with contributions of all authors. All authors have given approval to the final version of the manuscript.

**Acknowledgements**
This material is based upon work supported by the NSF/EPSCoR RII Track-1: Emergent Quantum Materials and Technologies (EQUATE) Award OIA-2044049, and NSF Award 1809800. I.F. acknowledges support from ERAF project 1.1.1.5/20/A/001. The research was performed in part in the Nebraska Nanoscale Facility: National Nanotechnology Coordinated Infrastructure and the Nebraska Center for Materials and Nanoscience (and/or NERCF), which are supported by NSF under Award ECCS: 2025298, and the Nebraska Research Initiative. We thank R. Mahbub for providing details on doing SEM on Fe-triazole nanoparticles, C. W. Kiat for patterning the diamond with reference letters by FIB, and M. Dowran for helping with the first Raman measurements.

## Supporting Information

### I. SEM measurements of Fe-triazole nanoparticles and nanorods

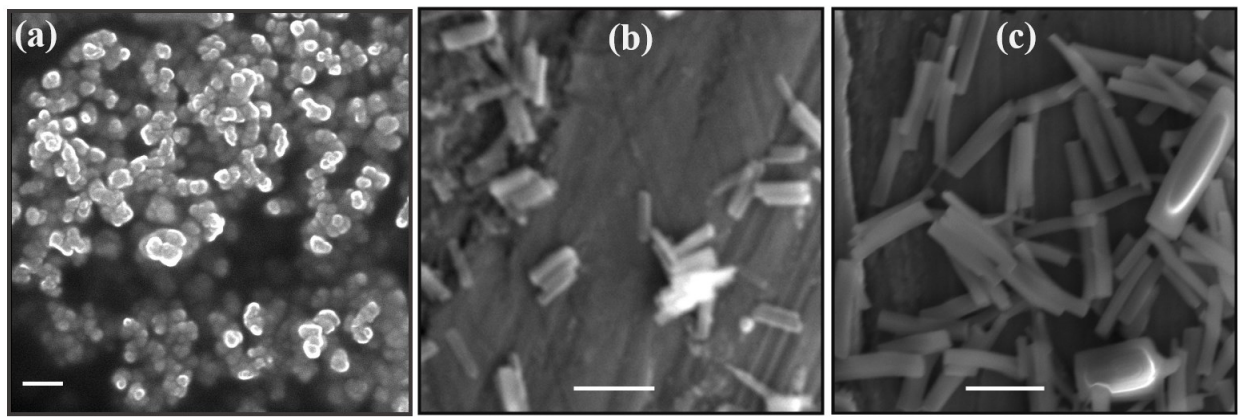

**Figure S1.** SEM images of the 20 nm **(a)**, 300 nm **(b),** and 1 μm **(c)** Fe-triazole nanorods. The scale bar is 100 nm in **(a)**, 500 nm in **(b)**, and 1 μm in **(c)**.

### II. Profile matching XRD analysis on measurements of Fe-triazole nanoparticles and nanorods.
XRD measurements performed on Fe-triazole 20 nm nanoparticles (**Fig. S2a**) and on 1 μm nanorods (**Fig. S2b**) are close to polymorph I and polymorph II respectively.[47] We performed profile matching by refining lattice parameter and peak shape to obtain lattice parameter and hence the density of the nanoparticles/nanorods (see the main manuscript).

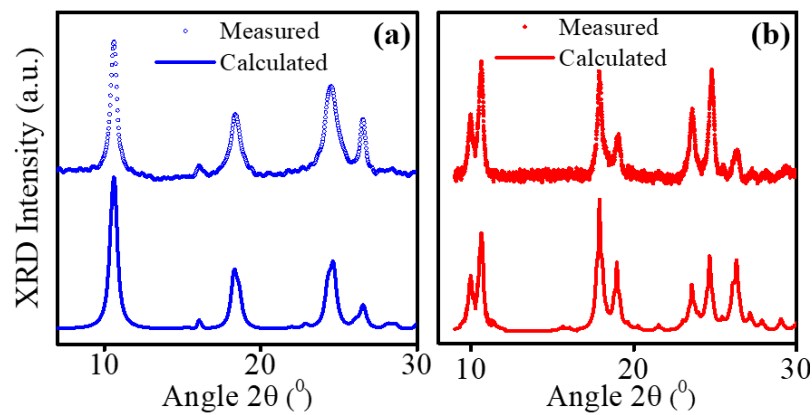

**Figure S2. (a)** Measured (open circles) and calculated (solid line) XRD spectra of the Fe-triazole 20 nm nanoparticles. **(b)** Measured (closed circles) and calculated (solid line) XRD spectra of the Fe-triazole 1 μm nanorods. The calculated curves are obtained by profile matching performing Fullprof software [47] and refining the lattice parameter and peak shape.

### III. Height measurements of the 1 μm Fe-triazole nanorods

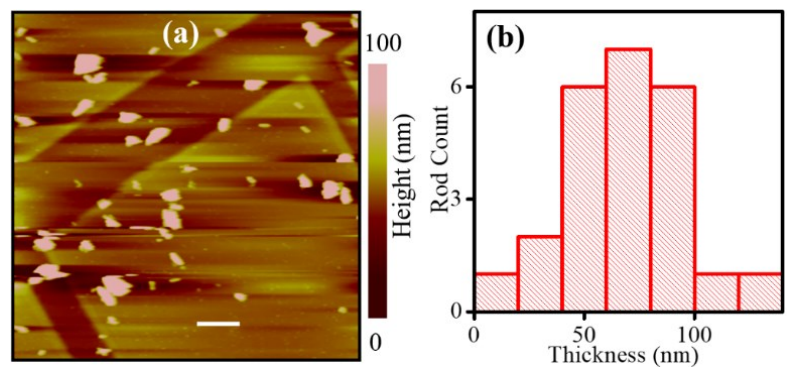

**Figure S3. (a)** AFM image of the 1 μm Fe-triazole nanorods deposited on the diamond substrate. **(b)** The height distribution of nanorods with mean height of ~75 nm ± 20 nm. The scale bar in **(a)** is 3 μm.

## IV. Magnetic thermal properties of Fe Triazole nanoparticle/nanorod powder

To confirm the temperature dependent spin states of the Fe-triazole nanorods and nanoparticles we used vibrating-sample magnetometer (VSM, Versal Lab 3T from Quantum Design) and measured the bulk magnetic molar susceptibility of the nanorods and nanoparticles powder as function of the temperature in the range of 298-400 K at an applied magnetic field of 2 T (temperature rate is 3.5 K/min). The VSM measurements in **Fig. S4a**, **Fig. S4b**, and **Fig. S4b** confirm the transition from the low spin (LS) state to the high spin (HS) at a temperature range of 370 K – 390 K for the Fe-triazole 20 nm nanoparticles, 300 nm nanorods, and 1 µm nanorods respectively. The VSM measurements in **Figure S4** are repeated over two cycles to remove any remnant solution in the clustered rods and to confirm a successful thermal hysteresis.

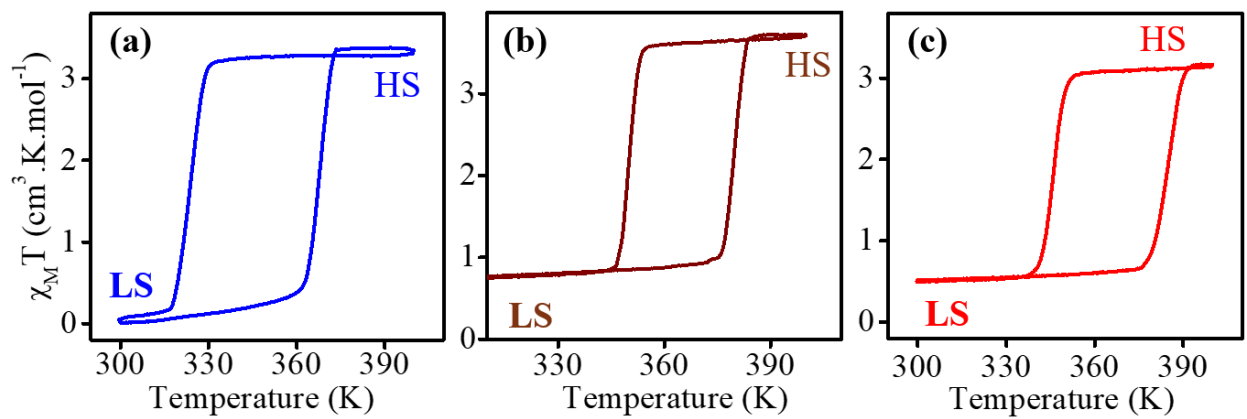

**Figure S4.** Magnetic molar susceptibility as function of temperature in the Fe-triazole 20 nm nanoparticles **(a)**, 300 nm nanorods **(b)**, and 1-µm nanorods **(c)**.

**Figure S5** shows a 4-cycle overlay comparison of the magnetic molar susceptibility as function of temperature for the Fe-triazole 20 nm nanoparticles. There is a difference between cycle 1 and cycle 2 confirming the remnant solution, however there is no change after cycle 2.

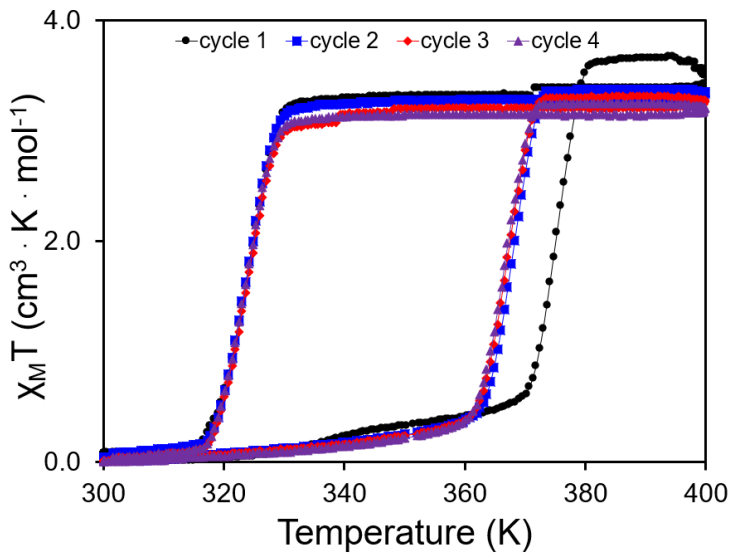

**Figure S5.** Magnetic molar susceptibility as function of temperature in the Fe-triazole 20 nm nanoparticles performed at 4 cycles (cycle 1: black circles, cycle 2: blue squares, cycle 3: red diamonds, cycle 4: purple triangles).

## V. Experimental setup: optical detected magnetic resonance

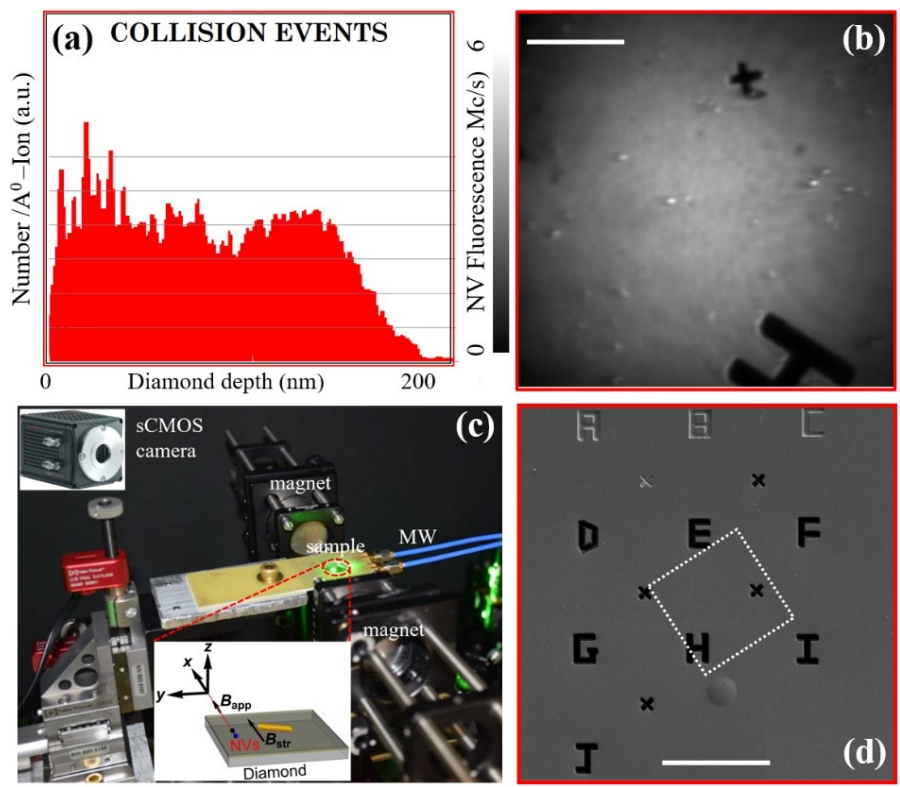

**Figure S6. (a)** SRIM vacancy-depth profile for the $^4$He$^+$ implantation in the diamond substrate (conditions are in the main text). **(b)** 39 μm x 39 μm NV fluorescence image of the scattered region in **(d)**, taken by the sCMOS camera at a laser intensity of 0.14 mW/μm$^2$. **(c)** ODMR microscope integrated with a wide-field sCMOS camera (upper insert) and a diamond substrate doped with a ~ 200 nm thick NV layer (lower insert). **(d)** A diamond substrate is patterned by FIB with reference letters for tracking the NV measurements. The scale bars in **(b)** and **(d)** are 10 μm and 39 μm respectively.

## VI. Raman and NV characterization of the LS state of the Fe-triazole 1 μm nanorods

For Raman measurements we used a green (532 nm) laser excitation, provided by Coherent Verdi V2 laser, to excite the sample via a Nikon 60x 0.7NA objective. The scattered reflected laser is then filtered by a Bragg band-pass filter and notch filter set before being sent into an isoplane320 spectrometer and captured with a PIXIS CCD.

To check whether the studied individual Fe-triazole nanorods preserved their LS magnetic properties after repetitive SEM and NV measurements we performed Raman measurements at room temperature (295 K) using the same experimental conditions in the main manuscript. **Figures S7a**, **S7b**, and **S7c** show Raman spectra in the wavenumber 100-340 cm$^{-1}$ for selected individual Fe-triazole 1 μm nanorods (SEM images are in the insert) measured at a laser intensity of 3.5 mW.μm$^{-2}$, way below the photoactivation effects induced by the 532-nm laser.[44] It is well studied that the Raman peaks in the LS appear at 132 cm$^{-1}$, 209 cm$^{-1}$ and 286 cm$^{-1}$, and are shifted towards lower wavelengths of 107 cm$^{-1}$, 139 cm$^{-1}$, and 184 cm$^{-1}$ when transitioning to HS upon heating.[53] This is related to Fe-N bond stretching which causes the change in vibrational frequencies.[53] Most of the Fe triazole nanorods that exhibit LS state generate a magnetic pattern in the optical detected magnetic resonance (ODMR) map (**Figs. S7d**, **S7e**, and **S7f**) with a dominated paramagnetic behavior.

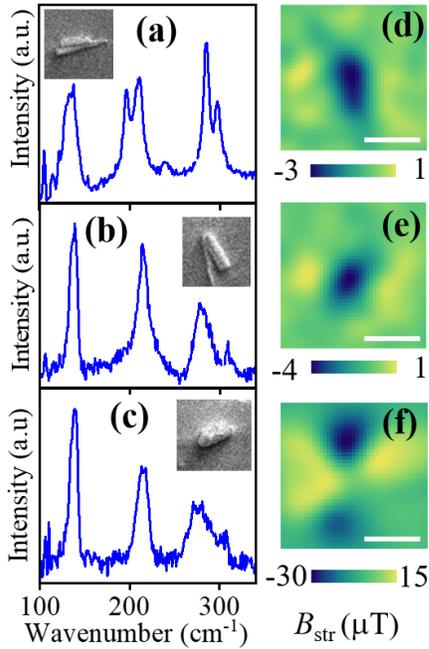

**Figure S7.** Normalized Raman spectrum of selected individual Fe triazole nanorods in **(a)**, **(b)**, and **(c)** with their corresponding NV OMDR maps in **(d)**, **(e)**, and **(f)** respectively for an applied field $B_{app}$= 250 mT. The sale bar is 1 μm. The SEM images of the nanorods are displayed in the upper inserts of **(a)**, **(b)**, and **(c)**. Raman and NV ODMR measurements confirm the LS state of the Fe Triazole nanorods that produce magnetic diploes with a stray field $B_{str}$ in the range of ± 30 μT.

## VII. Calculation of the magnetic stray field generated from the Fe-triazole nanorods

The Fe-triazole nanorods are magnetized under a magnetic field applied along the *x* direction ([111] NV axis) of the (110) diamond. To calculate stray fields produced from each nanorods we considered them as three-dimensional isotropic ellipsoids (**Fig. S8b**) made of a uniform magnetic material with magnetic susceptibility inferred from VSM measurements ($\chi \sim 1.17 \times 10^{-4} \pm 1.78 \times 10^{-5}$). The lateral dimensions of the Fe-triazole nanorods are inferred from SEM measurements (**Fig. S8a**) and their height is obtained from the AFM measurements in **Fig. S3**. For our magnetostatic finite-elements modeling we only considered the stray-field component along *x*-axis, with the nanorods magnetized along the same *x*- direction by the external field $B_{app}$. For accurate representation of the magnetic dipoles produced by the Fe-triazole nanorods, we integrated $B_{str}$ over an NV sensing depth of ~ 200 nm (**Fig. S6a**).

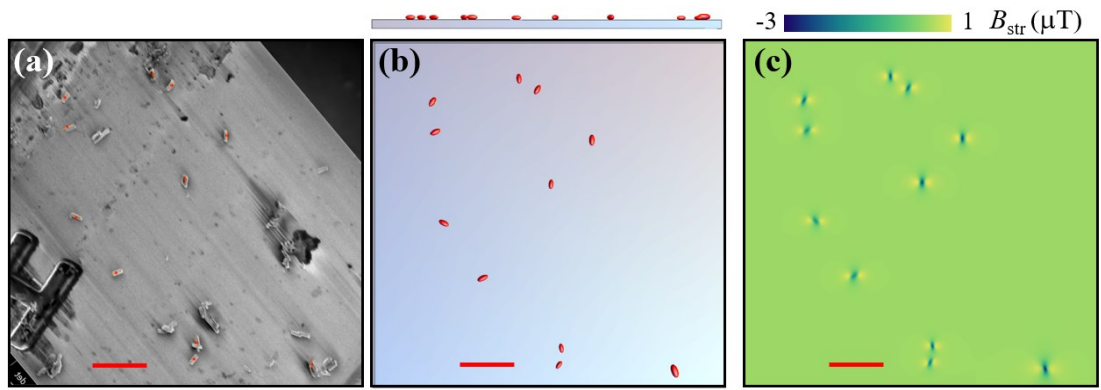

**Figure S8. (a)** SEM image of 1 μm Fe-triazole nanorods drop-casted on the diamond substrate. **(b)** 3D ellipsoids with dimension inferred from SEM are used to calculate the magnetic-field patterns. **(c)** Calculated magnetic-field pattern map from the 3D ellipsoids with a volume susceptibility $1.64 \times 10^{-4}$ and NV layer depth of 200 nm. The scale bar in **(a)**, **(b)**, and **(c)** is 10 μm.

The simulated stray fields $B_{str}$ generated by 1 μm Fe-triazole nanorods are shown in **Fig. S8c** at the volume susceptibility of $\chi \sim 1.64 \times 10^{-4} \pm 1.77 \times 10^{-5}$, which well agrees with NV and VSM measurements. We see a variation of $\chi$ values in the range of 30% depending on the nanorod orientation and purity (see section IX for further discussion).

**VIII. Additional NV measurements on Fe-triazole nanoparticles**

We correlated the optical image with NV ODMR map (measured after heating to 395 K and cooling to 295K) on Fe-triazole nanoparticles to check whether all clusters produce a magnetic pattern. The solid white squares in **Fig. S9a** show the Fe-triazole nanoparticle clusters that do not generate any magnetic patterns in the NV ODMR map (**Fig. S9b**) and can be either in the LS state with a diamagnetic behavior, and/or residue from drop casting the Fe-triazole/ethanol solutions. The difficulty in knowing precisely the type of clusters comes from the limited optical resolution of our ODMR microscope (~ 390 nm). We also used SEM (much better spatial resolution) to image the Fe-triazole clusters, but the significant charging issues made it difficult to get clearer images. Future measurements by coating the Fe-triazole nanoparticles on top of the diamond with a conductive polymer could help imaging individual nanoparticles below 20 nm.[65]

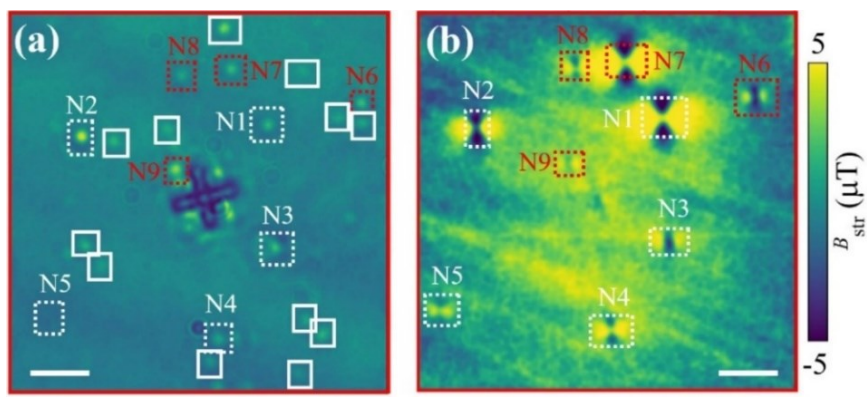

**Figure S9.** (a) Optical image of drop-casted Fe-triazole nanoparticles on top of the diamond. (b) NV magnetic-field map of all clusters in (a), measured after heating to 395 K and cooling to 295K. The scale bar in (a) and (b) is 5 μm.

The white scattered squares (N1-N5) are the Fe-triazole clusters that generated the magnetic patterns at 295 K (not heated), and the red scattered squares (N6-N9) are the clusters that generated magnetic patterns after heating to 395 K and cooling back to 295 K. We discussed the magnetic properties of these nanoparticles in the main manuscript.

**IX. Uncertainty in susceptibility measurements**

We performed NV measurements on the 300 nm Fe-triazole nanorods (**Fig. S10a**) at an applied magnetic field of 180 mT, and found of the 8 nanorods, 7 nanorods (dashed squares) produce weak magnetic stray field (≤ 7 μT), and 1 rod (solid squares) generate strong magnetic features ($B_{str} > 10$ μT). To further investigate the magnetic properties of the LS state at the individual level, we labelled selected Fe-triazole nanorods with numbers (S1-S8) in **Fig. S10a** and measured their amplitude $\Delta B$ magnetic stray fields as function of applied magnetic field $\underline{B}_{app}$ (**Fig. S10b**, **Fig. S14**). Most of the rods (e.g., S1, S2, S6, and S7) have a paramagnetic behavior with a susceptibility $\chi$ in the range of $1.6 \times 10^{-4}$ to $3.5 \times 10^{-4}$.

From NV measurements on the 1 μm Fe-triazole (**Figs. 4**, **S12**), 300 nm nanorods (**Figs. S10b**, **S14**), 20 nm nanoparticles (**Figs. 5**, **S10c**, **S13**) we found a variation of the slope of their stray field $\Delta B$ vs applied magnetic field $B_{app}$, which leads to an uncertainty in estimating $\chi$ within a range of 30 - 48% (see Section X). Similar variations were observed in individual hemozoin biocrystals, explained by the NV-layer imperfect distribution, image blur, and hemozoin crystal dimensions and shape.[43] As seen in **Fig. 4a** and **Fig. S1c** the 1 μm Fe-triazole nanorods have different shapes and are oriented along different angles within the applied magnetic field $B_{app}$. The uncertainty in determining χ in the nanorods/nanoparticles can be also explained by the anisotropic components of the magnetization along *x* and *y* directions,[43] as confirmed in the SEM images (**Figs. S1b**, **S1c**, i.e., elongation of the rods). In our magnetostatic model we considered a uniform isotropic medium to calculate the magnetic patterns. Usually, the magnetic anisotropy is significant in ferromagnetic/ferrimagnetic dots[55-57] and nanoparticles.[54]

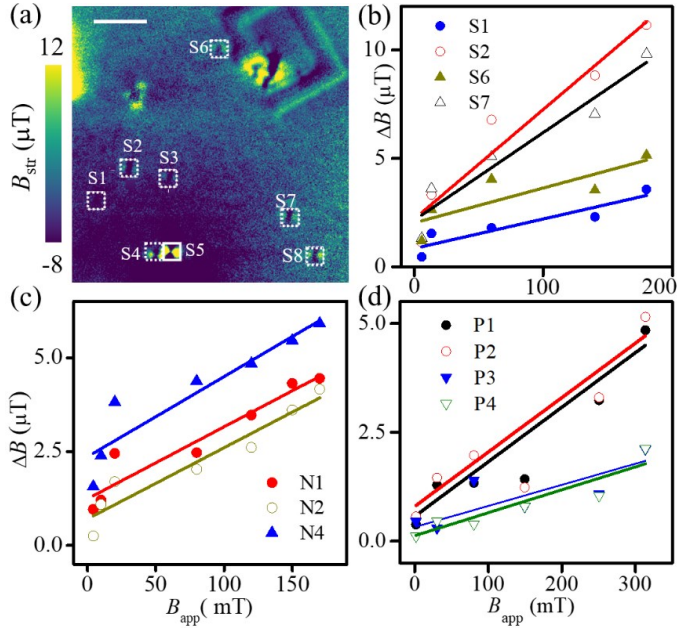

**Figure S10**: **(a)** NV ODMR map of the 300 nm Fe-triazole nanorods for $B_{app}$= 180 mT. The scale bare is 5 μm. Stray-field NV measurements on selected individual Fe-triazole 300 nm rods **(b)**, 20 nm nanoparticle clusters **(c)**, and 1 μm nanorods **(d)**.

## X. Summary of Δ*B*-$B_{app}$ measurements on Fe-triazole nanoparticles/ nanorods

Figure **S11a** displays the NV produced magnetic-field map of the Fe-triazole 1 μm rods in **Fig. 3a** at an applied magnetic field of 85 mT, labeled with particles P1-P33. The red marked rods (dashed/solid squares) are studied in the main manuscript, **Fig. 4**. **Fig. S11b** shows the histogram of the occurrence (number of nanorods) as function of magnetic pattern amplitude Δ*B*. Most of the particles produce a stary field < 10 μT.

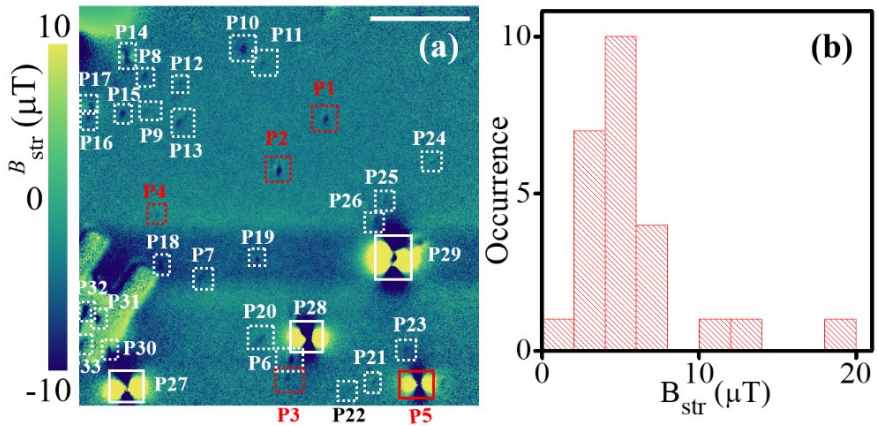

**Figure S11:** **(a)** NV magnetic-field map of all (33) 1 μm nanorods for $B_{app}$= 85 mT. The labeled rods with red numbers are studied in detail in **Fig. 4**. White scattered and solid squares are studied in detail in **Fig. S12**. Scale bare is 10 μm. **(b)** Histogram of the occurrence (number of nanorods) as function of the magnetic pattern amplitude Δ*B*.

**Figure S12** displays the magnetic pattern amplitude $\Delta B$ of the 1 μm nanorods P1-P33 as function of $B_{app}$. Linear dependence is observed in nanorods P1-P4, P6-P26, and P30-P33 which is typical of paramagnetic behavior. We used a linear function ($\Delta B = A\, B_{app}$) to fit the measurements (solid lines in **Fig. S12**), where $A$ is the curve slope, and it corresponds to the magnetic susceptibility $\chi$ of the nanorods. Susceptibility $\chi$ varies in the range from $1.65 \times 10^{-4}$ to $3.2 \times 10^{-4}$ in the 29 nanorods, which corresponds to a variation of 48%. For nanorods P5 and P27-P29 $\Delta B$ increases nonlinearly with applied magnetic field and saturates at $B_{app} \sim 100$ mT. This is typical superparamagnetic behavior and the measured $\Delta B(B_{app})$ is fitted well with a Langevin.[58]

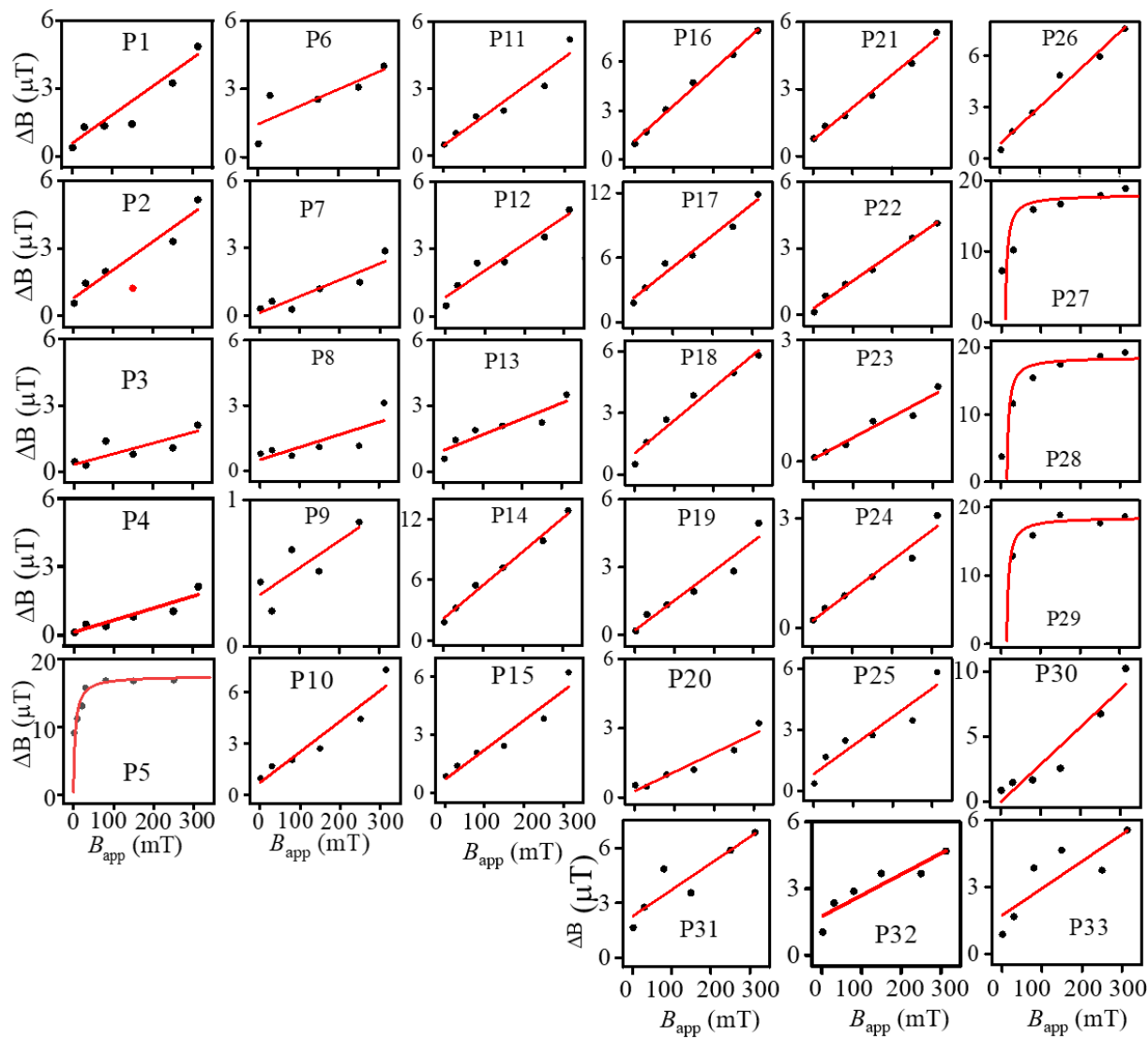

**Figure S12:** $\Delta B$-$B_{app}$ dependence for labelled (in **Fig. S11a**) 1 μm Fe-triazole nanorods (scattered closed circles: NV measurements, solid lines: fits). P1-P4, P6-33 are paramagnetic based on the linear fit. P5 and P27-P29 exhibit a superparamagnetic behavior and are fitted with a Langevin showing a saturation at ~ 100 mT

**Figure S13** shows the magnetic pattern amplitude $\Delta B$ of the 20 nm nanoparticle clusters N1-N9 as function of $B_{app}$. Linear dependence is observed which is typical of paramagnetic behavior.

The susceptibility χ varies in the range from 1.51 x $10^{-4}$ to 2.1 x $10^{-4}$ in the 9 nanoparticle clusters, which corresponds to a variation of 28%.

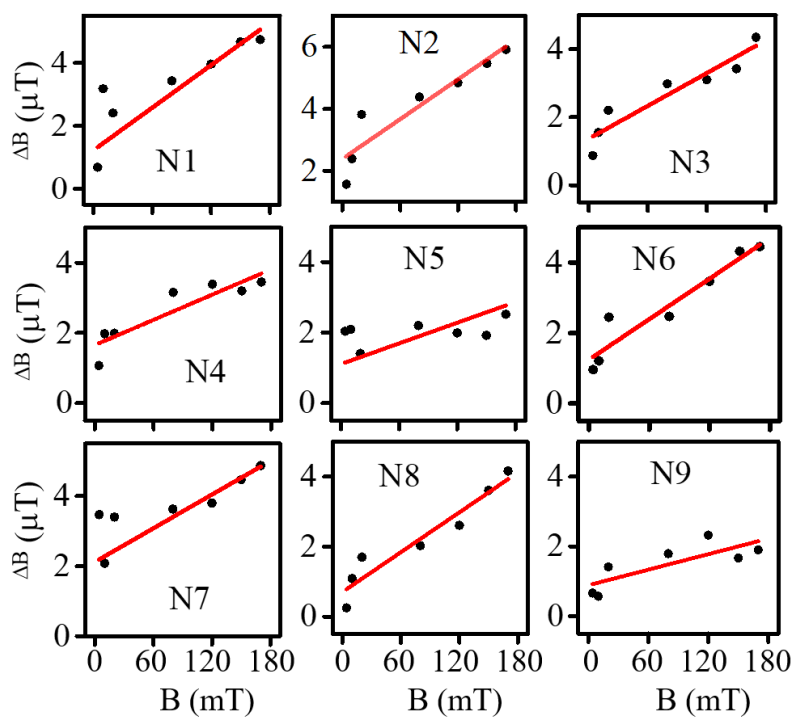

**Figure S13:** $\Delta B$ dependences on the $B_{app}$ for labelled (in **Fig. S9**) 20 nm Fe-triazole nanoparticle clusters (scattered closed circles: NV measurements, solid lines: fits). N1-N9 are paramagnetic based on the linear fit.

**Figure S14** shows the magnetic pattern amplitude $\Delta B$ of the 300 nm Fe-triazole nanorods S1-S8 as function of $B_{app}$. Linear dependence is observed which is typical of paramagnetic behavior. The susceptibility χ varies in the range from 1.71 x $10^{-4}$ to 4.11 x $10^{-4}$ in the 8 nanorods, which corresponds to a variation of 38%.

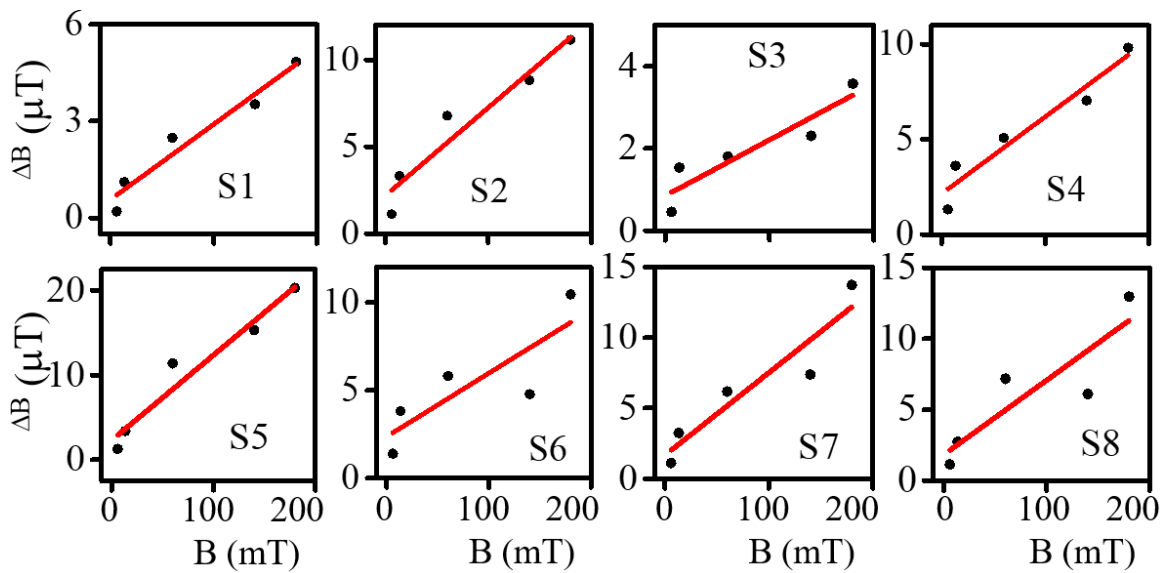

**Figure S14:** $\Delta B$ dependences on the $B_{app}$ for labelled (in **Fig. S10a**) 300 nm nanorods (scattered closed circles: NV measurements, solid lines: fits). S1-S8 are paramagnetic based on the linear fit.